\documentclass[conference]{IEEEtran}
\usepackage{amsmath}
\usepackage{cite}
\usepackage{pdflscape} 
\usepackage{booktabs}
\usepackage{siunitx}
\usepackage{tikz}
\usetikzlibrary{positioning, shadows.blur}
\IEEEoverridecommandlockouts
\usepackage{cite}
\usepackage{amsmath,amssymb,amsfonts}
\usepackage{algorithmic}
\usepackage{textcomp}
\def\BibTeX{{\rm B\kern-.05em{\sc i\kern-.025em b}\kern-.08em
    T\kern-.1667em\lower.7ex\hbox{E}\kern-.125emX}}
\usetikzlibrary{arrows.meta, positioning, shapes.geometric}
\usepackage{tikz}

\tikzstyle{conv} = [rectangle, draw, fill=blue!50]
\tikzstyle{fc} = [rectangle, draw, fill=green!50]
\tikzstyle{arrow} = [thick,->,>=stealth]
\usepackage{import}
\usetikzlibrary{quotes,arrows.meta}
\usetikzlibrary{positioning}

\def\edgecolor{rgb:blue,4;red,1;green,4;black,3}
\newcommand{\midarrow}{\tikz \draw[-Stealth,line width =0.8mm,draw=\edgecolor] (-0.3,0) -- ++(0.3,0);}

\usepackage{Ball}
\usepackage{Box}
\usepackage{RightBandedBox}
\usepackage{RightBandedBox2}

\def\FcColor{rgb:yellow,5;red,2.5;white,5}
\def\FcReluColor{rgb:yellow,5;red,5;white,5}
\def\DropoutColor{rgb:red,1;black,0.3}
\def\ReluColor{rgb:blue,5;green,2.5;white,5}
\def\SoftmaxColor{rgb:magenta,5;black,7}
\def\OutputColor{rgb:blue,5;green,2.5;white,5}

\usepackage{amsmath}
\usepackage{amsfonts}
\usepackage{caption}
\usepackage{subcaption}
\usepackage{multirow}
\usepackage{graphicx}
\usepackage{booktabs}
\usepackage{tabularx}
\usepackage{listings}
\usepackage{hyperref}
\makeatletter
\def\section{\@startsection{section}{1}{\z@}{-2ex plus -1ex minus -.2ex}{1ex plus .2ex}{\normalsize\centering\scshape\bfseries}}
\def\subsection{\@startsection{subsection}{2}{\z@}{-1.5ex plus -1ex minus -.2ex}{0.8ex plus .2ex}{\normalsize\scshape\bfseries}}
\def\subsubsection{\@startsection{subsubsection}{3}{\z@}{-1ex plus -1ex minus .2ex}{1ex plus .2ex}{\normalsize\itshape\bfseries}}

\usepackage{url}

\begin{document}
\title{Detection of Deepfake Environmental Audio}

\author{
    \IEEEauthorblockN{Hafsa Ouajdi }
    \IEEEauthorblockA{
    \textit{École Centrale Nantes} \\
    Nantes, France \\
    hafsaouajdimp@gmail.com}
    \and
    \IEEEauthorblockN{Oussama Hadder }
    \IEEEauthorblockA{ 
    \textit{École Centrale Nantes} \\
    Nantes, France \\
    natsuhadder001@gmail.com}
    \and
       \IEEEauthorblockN{Modan Tailleur, Mathieu Lagrange}
    \IEEEauthorblockA{\textit{Nantes Université,} \\
    \textit{École Centrale Nantes, } \\
    \textit{CNRS, LS2N, UMR 6004,} \\
     F-44000 Nantes, France \\
    first.last@ls2n.fr}
    \and
    \IEEEauthorblockN{Laurie M. Heller}
    \IEEEauthorblockA{\textit{Dept. of Psychology,} \\
    \textit{Carnegie Mellon University,} \\
    Pittsburgh, PA, U.S. \\
    hellerl@andrew.cmu.edu}
}

\markboth{Nov~2023}%
{Shell \MakeLowercase{\textit{et al.}}: Bare Demo of IEEEtran.cls for Computer Society Journals}

\maketitle
\begin{abstract}
With the ever-rising quality of deep generative models, it is increasingly important to be able to discern whether the audio data at hand have been recorded or synthesized. Although the detection of fake speech signals has been studied extensively, this is not the case for the detection of fake environmental audio. We propose a simple and efficient pipeline for detecting fake environmental sounds based on the CLAP audio embedding. We evaluate this detector using audio data from the 2023 DCASE challenge task on Foley sound synthesis.

Our experiments show that fake sounds generated by 44 state-of-the-art synthesizers can be detected on average with 98\% accuracy. We show that using an audio embedding trained specifically on environmental audio is beneficial over a standard VGGish one as it provides a 10\% increase in detection performance. The sounds misclassified by the detector were tested in an experiment on human listeners who showed modest accuracy with nonfake sounds, suggesting there may be unexploited audible features.  

\end{abstract}


\section{Introduction}

The rapid evolution of generative models, 
such as those based on Diffusion models, ushers in an era where the boundaries between reality and synthetic content tend to blur more and more. In audio synthesis, there exists a sizable literature dedicated to the detection of deep fakes in speech, aiming to detect adversarial attacks ranging from misinformation dissemination to identity theft \cite{EURECOM+6851}.

As generative models become more sophisticated, particularly those rooted in deep learning architectures, their capacity to produce eerily realistic audio forgeries has grown exponentially. Innovations such as Variational Autoencoders (VAEs) and Generative Adversarial Networks (GANs) have empowered malicious actors to craft audio content that is virtually indistinguishable from genuine recordings. 
Furthermore, deepfake audio is often used in conjunction with deepfake video to create more realistic and convincing fake movies. In light of these trends, the design of effective fake audio detection systems is important. 

Deepfake audio refers to audio that has been generated or augmented using deep learning techniques \cite{yi2023audio}. There are different types of Deepfake Audio: Text-to-Speech \cite{wu2015spoofing}, Voice conversion \cite{das2021towards}, emotion fakes\cite{zhao2022emofake}, scene fakes\cite{yi2022scenefake}, and partial fakes \cite{yi2021half}.

While deepfake detection for speech is well studied,  there appears to be little research on the detection of fake content for environmental sounds. This paper aims to fill this gap with respect to Foley sound synthesis, \textit{i.e.,} sound categories such as \textit{hand clap}, \textit{rain}, etc.

We propose a simple and effective fake detection pipeline based on CLAP embeddings \cite{elizalde2023clap}. Experiments are based on a publicly available audio dataset developed during Task 7 of the IEEE AASP DCASE 2023 Challenge \cite{choi2023foley}. This is a dataset of more than 6 hours of recorded audio and 28 hours of generated audio.

The paper is organized as follows. Section \ref{sota} gives a brief review of the relevant state of the art on deepfake detection in audio. Section \ref{approach} introduces the proposed deepfake detector which is benchmarked using an experimental protocol described in Section \ref{protocol}. Performance is discussed in Section \ref{res} and Section \ref{discussion} discusses the outcome of listening tests on detection mistakes of the proposed detector and directions for improvement. Code, supplementary material, and audio examples are available on the companion page.\footnote{Companion page: \url{https://mathieulagrange.github.io/audioFoleyDeepFake}}

\section{Related Work}
\label{sota}
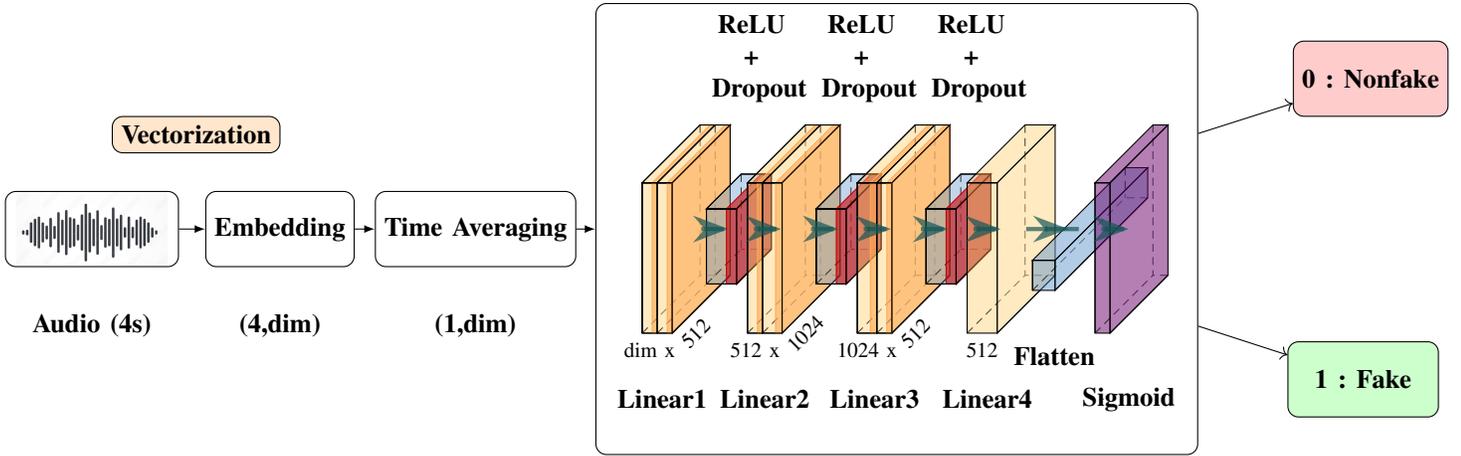
\begin{figure*}[h]
\centering
\begin{tikzpicture}
\tikzstyle{block} = [draw, fill=blue!20, rectangle, minimum height=3em, minimum width=6em, blur shadow={shadow blur steps=5}]
\tikzstyle{overlapped} = [draw, fill=blue!20, rectangle, minimum height=2.5em, minimum width=5em]
\tikzstyle{connection}=[ultra thick,every node/.style={sloped,allow upside down},draw=\edgecolor,opacity=0.7]
\node[draw, rounded corners, minimum width=2.3cm, minimum height=1cm] (audio) at (-20,0) {Audio};
\node[below = 0.5cm of audio] (Audio signal) {\textbf{Audio (4s)}};

\node[draw, rounded corners, minimum width=1cm, minimum height=1cm] (embedding) at (-17.5,0) {\textbf{Embedding}};
\node[below = 0.5cm of embedding] (caption) {\textbf{(4,dim)}};

\node[draw, rounded corners, minimum width=1cm, minimum height=1cm] (average) at (-14.9,0) {\textbf{Time Averaging}};
\node[below = 0.5cm of average] (caption) {\textbf{(1,dim)}};
\node[draw, rounded corners, fill = orange!20,above left = 0.5cm and -1cm of embedding] (caption) {\textbf{Vectorization}};

\node[fill=white!20, minimum width=7.3cm, minimum height=6cm] (boxall1) at (-9.2,0.3) {} ;

\node[fill=white!20, minimum width=7.3cm, minimum height=6cm] (boxall2) at (-9.3,0.2) {} ;
\node[fill=white!20, minimum width=7.3cm, minimum height=6cm] (boxall3) at (-9.4,0.1) {} ;
\node[draw, rounded corners, fill=white!20, minimum width=8.cm, minimum height=6cm] (boxall4) at (-9.3,0) {} ;

\node[right = 0.2cm of boxall1] (Audio signal) {\textbf{}};

\node[draw,rounded corners,fill=red!20, minimum width=1cm, minimum height=1cm] (fake box) at (-3,2) {\textbf{0 : Nonfake}} ;

\node[draw,rounded corners,fill=green!20, minimum width=2cm, minimum height=1cm] (fake box2) at (-3.1,-2) {\textbf{1 : Fake}} ;

\draw[-latex] (audio) -- (embedding) node[midway, above] {};
\draw[-latex] (embedding) -- (average) node[midway, above]{} ;
\draw[-latex] (average) -- (boxall4) node[midway, above]{} ;
\draw[->] (boxall4) -- (fake box);
\draw[->] (boxall4) -- (fake box2);
\node at (-20,0) {\includegraphics[width=2cm]{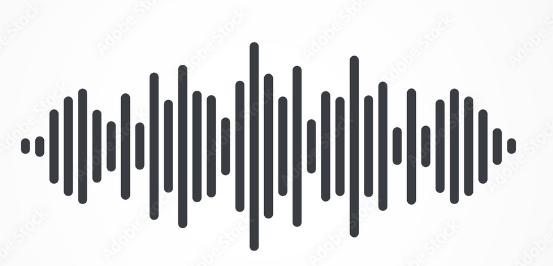}};
\pic[shift={(0,0,0)}] at (-12.3,0,0) {RightBandedBox={name=l1,caption=Linear1,%
        xlabel={{"dim x",""}},zlabel={512},fill=\FcColor,bandfill=\FcReluColor,%
        height=10,width={1,1},depth=10}};
\pic[shift={(0.3,0,0)}] at (l1-east) {RightBandedBox2={name=rd1,caption=ReLU + Dropout,%
        xlabel={{"",""}},zlabel=,fill=\ReluColor,bandfill=\DropoutColor,%
        height=5,width={2},depth=6}};
\pic[shift={(0.3,0,0)}] at (rd1-east) {RightBandedBox={name=l2,caption=Linear2,%
        xlabel={{"512 x",""}},zlabel={1024},fill=\FcColor,bandfill=\FcReluColor,%
        height=10,width={1,1.3},depth=10}};
\pic[shift={(0.3,0,0)}] at (l2-east) {RightBandedBox2={name=rd2,caption=ReLU + Dropout,%
        xlabel={{"",""}},zlabel=,fill=\ReluColor,bandfill=\DropoutColor,%
        height=5,width={2},depth=6}};
\pic[shift={(0.3,0,0)}] at (rd2-east) {RightBandedBox={name=l3,caption=Linear3,%
        xlabel={{"1024 x",""}},zlabel={512},fill=\FcColor,bandfill=\FcReluColor,%
        height=10,width={1.3,1},depth=10}};
\pic[shift={(0.3,0,0)}] at (l3-east) {RightBandedBox2={name=rd3,caption=ReLU + Dropout,%
        xlabel={{"",""}},zlabel=,fill=\ReluColor,bandfill=\DropoutColor,%
        height=5,width={2},depth=6}};
\pic[shift={(0.3,0,0)}] at (rd3-east) {Box={name=l4,caption=Linear4,%
        xlabel={{"512","output\_size"}},zlabel=,fill=\FcColor,%
        height=10,width=2,depth=10}};
\pic[shift={(0.7,0,0)}] at (l4-east) {Box={name=flatten,caption=Flatten,%
        fill=\OutputColor,opacity=0.5,height=2,width={1.5},depth=16}};
\pic[shift={(0.3,0,0)}] at (flatten-east) {Box={name=sigmoid,caption=Sigmoid,%
        fill=\SoftmaxColor,opacity=0.5,height=10,width=1,depth=10}};

\draw [connection]  (l1-east)    -- node {\midarrow} (rd1-west);
\draw [connection]  (rd1-east)   -- node {\midarrow} (l2-west);
\draw [connection]  (l2-east)    -- node {\midarrow} (rd2-west);
\draw [connection]  (rd2-east)   -- node {\midarrow} (l3-west);
\draw [connection]  (l3-east)    -- node {\midarrow} (rd3-west);
\draw [connection]  (rd3-east)   -- node {\midarrow} (l4-west);
\draw [connection]  (l4-east)    -- node {\midarrow} (flatten-west);
\draw [connection]  (flatten-east)    -- node {\midarrow} (sigmoid-west);

\end{tikzpicture}
\caption{Overview of the pipeline used in the experiments for the Deepfake detection, with a representation of the MLP's network architecture. The value of \textit{dim} depends on the embedding method used. }
\label{architecture}

\end{figure*}

This section discusses machine-learning techniques for Deepfake Audio detection. In \cite{florez2020applied}, the authors use linear regression to detect fake voices. The features are extracted using a signal's entropy calculated based on Shannon's equation for each second of the signal and for the whole signal. The model reaches 98\% accuracy, with all fake audio being correctly detected. In \cite{singh2021detection}, the authors adopt a combination of Bispectral analysis and  MFCC  (Mel Frequency Cepstral Coefficients) for speech detection and feature extractions. The former helps to identify components in generated audio that are not present in human speech and the latter is used to detect the features in human speech tied to the vocal tract that are absent in the AI-generated speech. The features are then fed into seven different  ML algorithms. The results show that the Quadratic SVM model gives the highest accuracy of 96.1\%. In \cite{borrelli2021synthetic}, the feature vector is extracted by filtering the signal under analysis and extracting different statistics with short-term and long-term prediction order. The features are then fed to a Linear SVM and Random Forest model. The results show that the SVM outperforms the RF in different conditions. \\
Traditional machine-learning models require manual feature extraction. Consequently, research was directed towards more sophisticated feature extraction algorithms based on deep learning models.
\\
In \cite{altalahin2023unmasking}, they employed both CNNs (Convolutional Neural Net), using local dependencies, and LSTMs (Long Short-Term Memory), using both local and sequential dependencies, to classify audio based on MFCCs. The former model helps achieve an accuracy of 80\%. A CNN has been used with a similar strategy in \cite{qais2022deepfake}, which compares different approaches (MFCC, STFT, FTT of the signal) to deduce that MFCC gives the best results and also that CNNs are beneficial in such tasks because they can learn to detect important features. In \cite{ballesteros2021deep4snet}, a shallow CNN architecture, Deep2Net, was used. The authors transformed the audio classification task into a computer vision problem by using a histogram of the audio. The model reached a global accuracy of 0.985.

\section{Proposed approach} \label{approach}

The detection of fake sounds is treated as a binary classification task.
As can be seen in Figure \ref{architecture}, the proposed architecture for solving this task leverages the power of pre-trained audio embeddings. This allows us to simplify the learning process by requiring less training data and using less power as the learnable part of the detector consists of three dense layers, as shown in Section \ref{architecture}. Each layer is followed by 1) ReLU activation functions to reduce the backpropagation errors and accelerate the learning process, and 2) dropout layers for regularization to avoid overfitting. The final dense layer is followed by a sigmoid function as the output activation function.

To provide input to the decision module, we tested 3 embeddings produced by deep learning architectures, namely: \textsc{vgg}, \textsc{clap}, and \textsc{pann}.

The \textsc{vgg} embedding is produced by the VGGish network, a convolutional neural network pre-trained for audio classification tasks, and is adapted from the VGG \cite{simonyan_very_2015} image classification architecture. This method adapts the VGG image classification network to the audio domain by converting audio signals into log-mel spectrograms, which are visual representations of the frequency and amplitude of the sound. The spectrograms are then fed into the network, which was pre-trained on a large-scale YouTube dataset with diverse audio categories, such as music, speech, or animal sounds. 
\\
The \textsc{clap} embedding is produced by the Contrastive Language-Audio Pretraining (Clap) model \cite{elizalde_clap_2023}. By leveraging a contrastive learning approach, CLAP enables models to learn representations that encode the semantic content of language and the acoustic characteristics of audio simultaneously. During pre-training, these models are trained using contrastive learning objectives to project both language and audio inputs into a shared embedding space. The model learns to encode similar language-audio pairs closely together while pushing dissimilar pairs apart. This process enables the extraction of rich embeddings that capture semantic and acoustic similarities between inputs.
\\
The \textsc{pann} embedding is produced by Pretrained Audio Neural Networks (PANNs) \cite{10.1109/TASLP.2020.3030497}, which are yet another class of models. These networks are trained on spectrogram representations of audio signals that capture both the frequency and temporal characteristics of sound. The pre-trained models are usually initialized with weights from models pre-trained on large-scale audio datasets, such as AudioSet or UrbanSound, which contain diverse audio categories ranging from musical genres to environmental sounds.

\section{Experiments}
\label{protocol}

\subsection{Dataset}

The \textbf{DCASE2023 Challenge} \cite{DBLP:journals/corr/abs-2305-15905} aimed to produce fake environmental sounds by training on nonfake sounds. It consisted of two tracks: Track A, with 10 generation systems from participating teams using varied Foley sound synthesis methods and permitting limited use of external resources; and Track B, with 28 generation systems based on provided code, prohibiting external resources. Both tracks accepted rule-based and ensemble systems, provided they utilized sounds only from the development set.

The resulting dataset consists of both nonfake and fake sounds across seven distinct sound classes (\textit{dog\_bark, footstep, gunshot, keyboard, moving\_motor\_vehicle, rain, and sneeze\_cough}). Specifically, the dataset comprises 5,550 nonfake sounds alongside 25,200 fake sounds, the latter of which were generated by challenge participants. These sounds are analyzed through the lens of four different embeddings, detailed in section II.B, each with its specific dimensionality: VGGish embeddings at (4,128), Microsoft CLAP (MS-Clap) embeddings at (4,1024), and two PANNs embeddings at (4,2048), where '4' refers to the four-second duration of the audio clips. For training and evaluation purposes, we separate the dataset into a training set (70\%), a validation set (10\%), and an evaluation set (20\%) \footnote{Dataset of recorded and generated audio: \url{https://zenodo.org/records/8091972}}.

\begin{table*}[h]
    \centering
    \begin{tabular}{c|cc|cc|cc|cc}
    \hline
    & \multicolumn{2}{c|}{VGGish} & \multicolumn{2}{c|}{ MS-Clap} & \multicolumn{2}{c|}{PANN\_Wavegram\_logmel } & \multicolumn{2}{c}{ PANN\_32k} \\ \hline
    \textbf{Predicted}  & \textbf{Nonfake (\%)} & \textbf{ Fake (\%)} & \textbf{ Nonfake (\%)} & \textbf{Fake (\%)} & \textbf{ Nonfake (\%)} & \textbf{Fake (\%)} & \textbf{ Nonfake (\%)} & \textbf{Fake (\%)} \\ \hline
    \textbf{Nonfake} & 15 & 3 & 17 & 1& 14 & 3& 14 & 5 \\
    \textbf{Fake} & 9 & 73 & 1 & 81& 4 & 79 & 3 & 77 \\ \hline
    \end{tabular}
    \caption{Confusion Matrix Evaluation for the three embeddings}
    \label{tab:confusion_matrix_combined}
\end{table*}

\subsection{Training Procedure}
 
Figure \ref{architecture} shows the pipeline used in the experiment protocol. As we have four different input embeddings, the procedure yielded four models to use and compare: VGGish, MS-Clap, PANN Wavegram, and PANNcnn14-32k. Note that we take the time average of each embedding 
to avoid the temporal sequence aspect (four-second duration) affecting the deepfake detection. 

These models were trained using the Adam optimizer, configured with a learning rate of 7e-4. The training protocol was standardized across models, with a batch size of 128 samples, a decision threshold of 0.5 on both training and validation predictions, and training over 100 epochs. Throughout this process, the Binary Cross-Entropy (BCE) loss function given in Equation \ref{eq:binary_cross_entropy} is used to optimize the weights parameters in the backpropagation process: 

\begin{equation}
J(y, \hat{y}) = - \left(y \cdot \log(\hat{y}) + (1 - y) \cdot \log(1 - \hat{y})\right)
\label{eq:binary_cross_entropy}
\end{equation}
Where $y$ is the true label (0 or 1) and $\hat{y}$ is the predicted probability.

During training, we implement a checkpointing mechanism that saves the state of the model every 10 epochs of training. Following the best-accuracy selection criterion \cite{Pham2020ProblemsAO}, the final model is chosen based on the checkpoint demonstrating the highest validation accuracy. For each implementation of the detector, 10 training runs are performed and statistics are reported. We carry out all the experiments on a machine with 4 physical cores and 8 threads, 16GB of RAM, and an NVIDIA GeForce RTX 3060 graphic card with 6GB of RAM. To address the imbalance between fake and nonfake samples, we employ a balancing technique on the training set.  

\section{Results} \label{res}

\subsection{Inference Time}

To evaluate the computational efficiency of each model, we assessed the time required for the embedding process and the model’s inference for each embedding on one random audio sample across 100 runs. The inference time (in seconds) is then computed by averaging over the 100 inference times. In Table II, the averaged inference time is expressed as a percentage of the real-time duration of the audio sample.  This metric provides a clear understanding of the model’s speed relative to the actual audio duration.

\subsection{Overall Accuracy}

We calculate both the relative overall validation and evaluation accuracy, along with the time per epoch in seconds (s), and generate a Confusion Matrix for the evaluation dataset. As presented in Table \ref{tab: Embeddings, networks, and training settings}, statistics such as Overall Evaluation Accuracy and Training Time per Epoch are detailed for each model. Among the experiments conducted under the same conditions, the  MS-Clap model outperformed others with the highest Evaluation Accuracy of 98.02\%, surpassing both PANN models by 5\% and the VGGish model by 10\%.

\begin{table}[h]
\centering
\caption{Statistics on Overall Accuracy and the Inference Time expressed as a real-time percentage.}
\begin{tabular}{@{}lccc@{}}
\toprule
{Model x Embeddings} & \multicolumn{1}{c}{Accuracy (\%)} & \multicolumn{2}{c}{Inference Time (\%)}  \\ \cmidrule(r){2-2} \cmidrule(lr){3-4}
& Mean  & Averaged percentage \\ \midrule
VGGish & 88.11 $\pm$ 0.73 &  0.423\\
MS-Clap & \textbf{98.02} $\pm$ 0.18 &  1.82 \\
PANN-Wavegram & 93.15 $\pm$ 0.34 &  0.318 \\
PANNcnn14\_32k & 93.04 $\pm$ 0.32 &  0.234 \\ \bottomrule
\end{tabular}
\label{tab: Embeddings, networks, and training settings}
\end{table}

Among the 40 identical runs, we pick the best-performing run for each model (\textit{i.e.}, the highest evaluation accuracy). The confusion matrices resulting from the four chosen models demonstrate a varied performance across the different architectures. The VGGish model shows difficulty with correctly classifying nonfake sounds, as it produces 9\% false positive judgements. The MS-Clap model shows the best proficiency in identifying the nonfake category of sounds, with the lowest false prediction rate (2\% of the dataset). On the other hand, the PANN-Wavegram and the PANN\_32k models behave approximately the same by having a high evaluation accuracy and also a low false prediction rate, yet they do not perform as well overall as the Clap model.

\subsection{Statistical Analysis}
We test the four models in the same configurations as shown in Table \ref{tab: Embeddings, networks, and training settings}. We train the models under each configuration for 10 iterations. We then perform the Mann-Whitney U-test to assess the similarity of accuracy distribution. The U-test is adopted instead of the t-test since it doesn't impose the assumption of normal distribution. In our case, the Null hypothesis assumes that run A is similar to run B. If the p-value $<$ 0.05, we can confirm that the alternative hypothesis is true (run A is similar to run B) with a confidence interval of 95\%.\\
The accuracy of MS-Clap is higher than that of Pann-Wavegram with 5.02\%, and that of PANN 32k with 5.11\%, and that of VGGish with 10.23\%. In all cases, the U-test confirms that those differences are significant, as for all pairs the Null hypothesis is rejected.

\begin{table*}[h]
\centering
\begin{tabular}{|c|c|c|c|c|c|c|c|}
\hline
\textbf{Sound Class} & \textbf{dog\_bark} & \textbf{footstep} & \textbf{gunshot}
& \textbf{keyboard} & \textbf{moving\_motor\_vehicule} & \textbf{rain} & \textbf{sneeze\_cough}  \\
\hline
\textbf{Accuracy (\%)} & 99  & 98.1 & 98.3 & 97.7 & 98.1 & 98.5 & 97.7\\
 \hline
\end{tabular}
\caption{Per Class Accuracy of the MS-Clap  deepfake predictor.}
\label{tab:MLP_accuracy_Class}
\end{table*}

\subsection{Class-Wise Accuracy}
As the MS-Clap model is the best-performing model among the four trained models, we chose to carry out our next analysis with this model. To have a better understanding of the behavior of the model's performance, we look closely at the model's accuracy for each class (Table \ref{tab:MLP_accuracy_Class}). This closer look helps us understand how the model performed with different types of data. We find that the model was consistently good at identifying all classes (98\% approximately for each class), with \textit{dog\_bark} being the class where the model performs the best, and it performs slightly worse than the others when it has to predict on \textit{keyboard} and \textit{sneeze\_cough} classes.



 
\subsection{Accuracy vs. Generator Quality}

We now compare our classifier's performance against the outcomes of the DCASE2023 challenge task 7. For this purpose, we conduct a comparison and analysis between the scores of our predicted likelihoods by using the BCE as a score function (the higher, the better), and the Fréchet Audio Distance (FAD) scores \cite{DBLP:journals/corr/abs-2305-15905} used for the official rankings of the challenge (the lower, the better). 





We observe a significant difference in correlation across the two tracks. While Track A demonstrated a strong negative correlation of -0.86, indicating a general agreement between the two scoring methods, Track B showed a notably lower correlation of -0.27. This discrepancy may be due to two major differences between the two tracks: Track B imposed restrictions on external resources, whereas Track A had none, and Track B had more systems than Track A (28 versus 10). Furthermore, the low correlation in Track B could be due to the diversity, complexity, and originality of the features used in the generating systems. For instance, feature extraction methods such as log-mel spectrogram and log-magnitude spectrogram were used by Track B's systems, while in Track A the majority used Gaussian latent variables and a spectrogram for feature extraction; additionally, some data augmentation strategies such as time masking, tanh distortion, and sound wrapping were used in Track B while being unused in Track A. 

\section{Listening Test}

\label{discussion}

Human listening suggests opportunities for improvement in the 
detection of distortion, realistic echoes and temporal structure, and patterns of noise and repetition which may be noticeable to humans but not to a frame-based detector such as the one considered in this study. 

\subsection{Data}

\subsubsection*{Incorrect Positives}

Erroneously classifying a nonfake sound as a fake is an "Incorrect Positive" whereas classifying a fake sound as a fake is a "Correct Positive". We examined 38 Incorrect Positive sounds, spanning all 7 categories, based on the erroneous high likelihood of the MS-Clap model. These sounds are from the DCASE development and evaluation dataset that were presumed to be real recordings; however, although these sounds were aurally screened by the organizers, the datasets may have nonetheless contained a few Foley or processed sounds. To detect this possibility and also to detect other reasons the system incorrectly classified a sound as fake, all authors carefully listened to these sounds. We noted two recordings that had very strong echoes. We suspected that two recordings were Foley-simulated footsteps rather than recordings of real footsteps. Some recordings included non-target sounds (e.g. footsteps during rain, or a snore after a sneeze). Many sounds had significant background noise. We verbally characterized five sound features that may have triggered an incorrect fake detection and used them to create a menu of "reasons" for a person suspecting that an audio is fake. 


\subsubsection*{Incorrect Negatives}

The 43 fake sounds that were erroneously classified as nonfake by the MS-Clap detector are referred to as "Incorrect Negatives". There were sounds from all 7 categories. We noted a couple of recordings that contained very little sound (perhaps a brief thud) and so there was not much available acoustic information. Many sounds had significant background noise, and often the noise sounded more like static than the typical noise heard in real recordings, such as environmental noise or distortion from overloading a microphone (clipping/saturation). Sometimes the onset of the target sound contained a burst of noise. One sound had audible repetition; evidence of splicing/editing was confirmed via a spectrogram. 


\subsection{Methods}

Twenty CMU undergraduates participated for academic credit (Mean age=19.8; 9 female, 10 male, 1 nonbinary) who reported normal hearing and passed a binaural screening survey to ensure careful listening over headphones. They used an online platform (gorilla.sc) and gave online consent approved by CMU’s IRB. Each sound was accompanied by its category name on the screen (e.g. “dog bark”), and each sound was played once before a judgement was made. All 7 sound categories from the System’s test set were represented in the set of 43 fake and 38 real sounds on which the ML system made errors. Participants were instructed to respond “real” if the sound seemed like a real recording of an event, and to respond “fake” if it seemed like it was generated by a computer program. They were told that the existence of background noise was not an indicator of fakeness and that cues could be present in either the sound from the category or the background noises.

\subsection{Outcomes}

When a fake (model-generated) sound was missed by the ML system (a system incorrect negative), the percent of sounds that listeners judged as fake (listener correct positives, M=49\%) was not significantly greater than chance, z=-0.07, p=0.6.  When a nonfake (recorded) sound was judged as fake by the ML system (system incorrect positive), the percent of sounds that listeners judged as real (listener correct negatives, M=71\%) was greater than chance, z=1.34, p $<$ 0.0001. This human sensitivity advantage implies that some acoustic features which indicated fakeness to the ML system did not seem artificial to human listeners.
When a sound was judged as fake by a participant, they indicated one of six possible reasons in the following proportions: 
The audio of the target sound is distorted (underwater, wobbly): 34.2\%
The sound does not sound like an example of the target category. 32.2\%
I hear suspicious artifacts in the background (whining, etc.): 12.9\%.
The sound is too brief or quiet to be able to give a reason. 10.0\%
The sound is too repetitive. 6.8\%
Other: 4\%.


\section{Conclusion}

We documented the first study of deep fake detection for environmental audio generation. The experiments reported considered a public dataset of sounds generated  by 10 state of the art systems and another public dataset of environmental audio recordings. 

The excellent results of the proposed deep fake detector demonstrate that, to this day, generated sounds can be detected rather easily. The listening test provide some interesting cues that can suggest research directions for the improvement of the generators.

\section{Acknowledgments}

The authors thank Justine Sullivan and Anjelica Ferguson for helping with the blind listening test described in Section \ref{discussion}.


\end{document}